\documentclass[11pt]{article}
\usepackage{bm}
\usepackage{geometry}                
\newcommand{\be}{\begin{equation}}
\newcommand{\ee}{\end{equation}}
\newcommand{\bea}{\begin{eqnarray}}
\newcommand{\eea}{\end{eqnarray}}

\usepackage{graphicx}
\usepackage{amssymb}
\usepackage{epstopdf}
\title{Transverse nucleon structure and multiparton interactions
\footnote{Presented at the LICracow School of Theoretical Physics "Soft side of the LHC", Zakopane, Poland, June 11-19, 2011.}}

\author{Mark Strikman\\
Penn State University,  Univeristy Park, PA 16802, U.S.A.
}
\date{}                                           

\begin{document}
\maketitle
\abstract{
The transverse structure of the nucleon as probed  in  hard exclusive processes plays  critical role in the understanding of the structure of the underlying event in  hard collisions  at the LHC, and multiparton interactions. We summarize results of our recent studies  of manifestation of transverse nucleon structure in the hard collisions at the LHC, new generalized parton distributions involved in multiparton interactions, presence of parton fluctuations. The kinematic range where interaction of fast partons of the projectile  with the  target reach black disk regime (BDR) strength is estimated. We demonstrate that in the BDR  postselection effect leads to effective fractional energy losses. This effect explains regularities of  the single and double  forward  pion production in $ dAu$ collisions at RHIC and impacts on the forward physics in $pp$ collisions at the LHC.
}

\section{Introduction}
The start of the LHC puts at a forefront the task of the describing the high energy proton - proton collision events in the whole their complexity. In particular, to search for new particles it is necessary to understand the structure of the underlying structure of  events with  dijets.  The knowledge of the 
  inclusive cross sections of hard binary collisions, which are expressed  through the convolution of the hadron parton distribution functions (PDF's) and the hard parton - parton interaction cross section, is not sufficient for these purposes.
 
 A natural framework for  description of the complete picture of the high energy interaction is the impact parameter representation of the collision. Indeed, 
in high--energy $pp$ scattering angular momentum conservation implies
that the impact parameter $b$ becomes a good quantum number, and it is 
natural to consider amplitudes and cross sections in the impact 
parameter representation. The nucleon's light--cone wave functions, 
describing the partonic structure at a low resolution scale, can be
expressed in terms of the longitudinal momentum fractions of the 
partons and their transverse positions relative to the 
center--of--momentum: $\psi_p(x_i,\rho_i)$. 

\begin{figure}[t]  
  \vspace*{-0.3cm}
   \hspace{2cm}\includegraphics[width=1.0\textwidth]{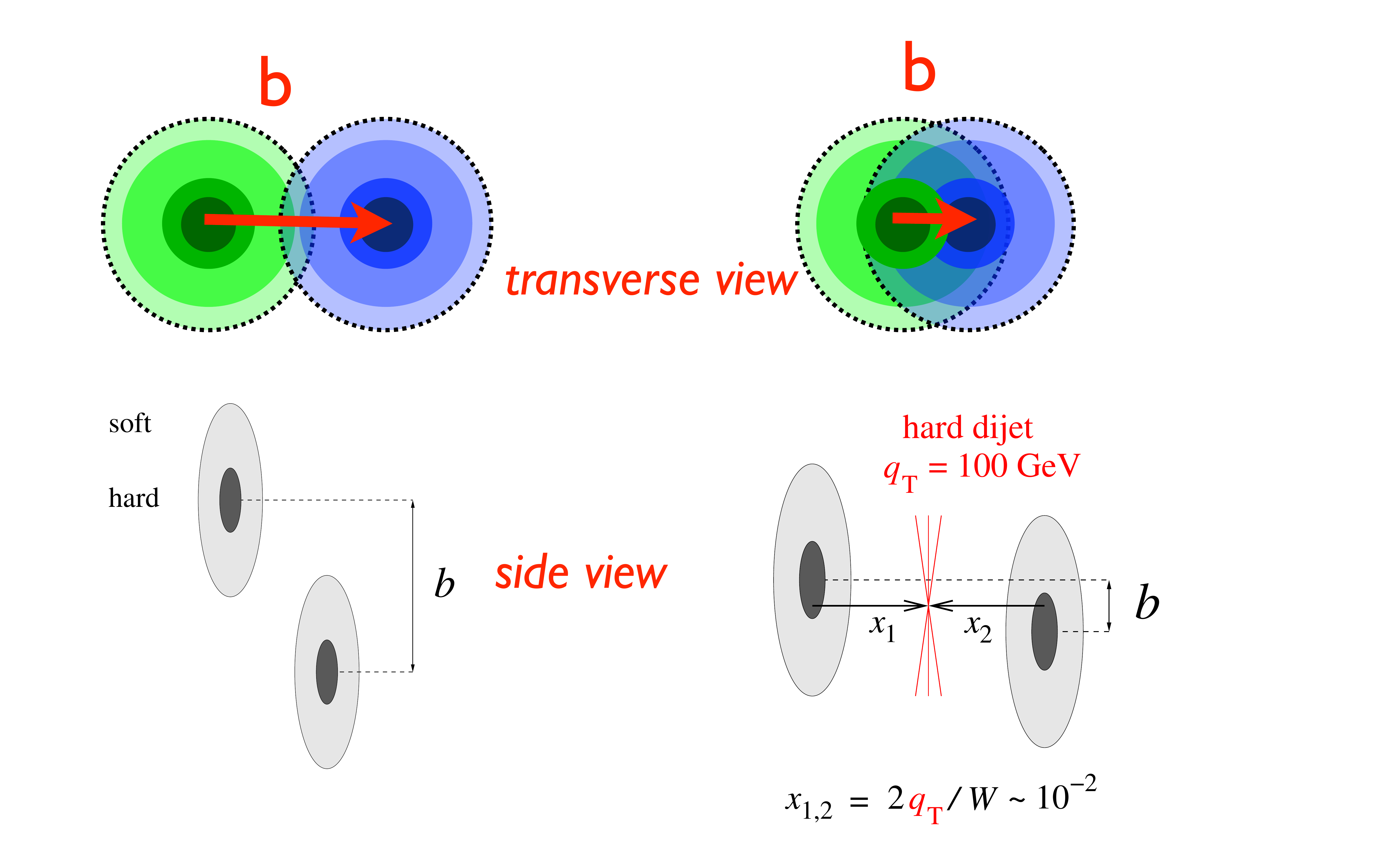} 
   \caption{$pp$ collisions at small and large impact parameters. 
  Transverse and side views. Dark (gray) disks  correspond to the areas occupied by hard (soft) partons.}
    \label{transverse}
 \end{figure}

The studies of exclusive hard processes lead to the conclusion that the transverse distribution of gluons with  $x\ge 10^{-3}$ than the distribution of soft partons involved in generic inelastic  $pp$ collisions (Sections 2 and 3). As a result,  
the events with a dijet trigger  should occur, in average, at much smaller impact parameters than the minimum--bias inelastic events - see 
Fig.~\ref{transverse}. Probability  of multiple soft and hard interactions is much higher for  head-on collisions than for peripheral collisions. Hence  one expects a much more active final states for the dijet triggered events than for the 
minimum-- bias events.
 
 To describe the transverse geometry of the $pp$ collisions with production of a dijet it is convenient to consider  probability to find a parton with given $x$ and transverse distance $\vec{\rho}$ from the nucleon center of mass, $f_i(x_i,\vec{\rho}_i)$. This quantity allows a formal operator definition, and it is referred to as  the diagonal generalized parton distribution(GPD). It is related to non-diagonal GPDs which enter in the description of the exclusive meson production (Sect. 2).
 The transverse geometry of the $pp$ collision with production of a dijet is  represented in 
 Fig.2.
 \begin{figure}[t]  
  \vspace*{-0.3cm}
   \hspace{2cm}\includegraphics[width=0.9\textwidth]{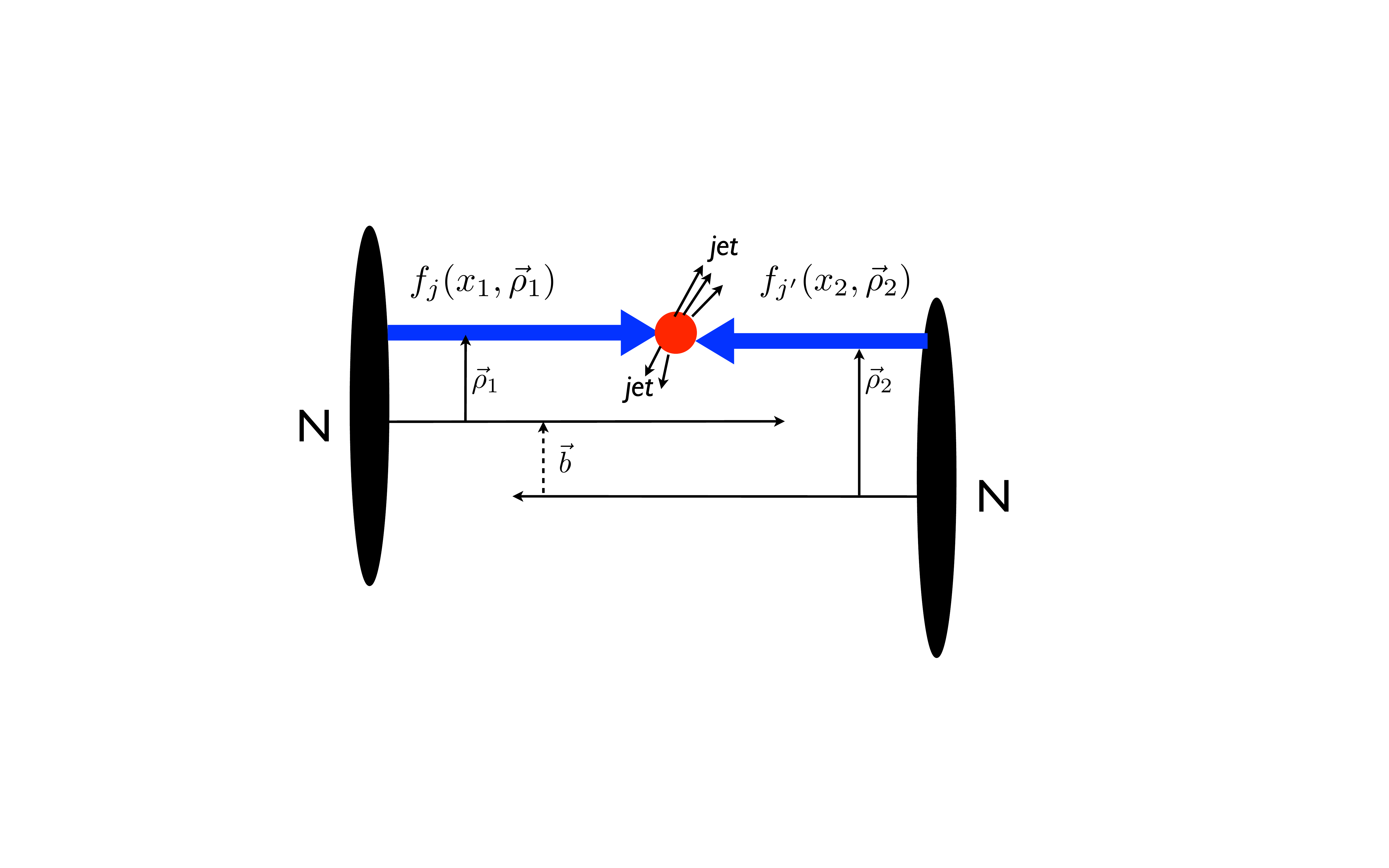} 
   \caption{Transverse geometry of hard collisions.}
    \label{geom}
 \end{figure}

 For inclusive cross section in the pQCD regime the  transverse structure does not matter - the cross section is expressed through the convolution of parton densities. Indeed, we can write 
 \bea
 \sigma_h \propto \int d^2bd^2\rho_1d^2\rho_2 \delta(\rho_1+b-\rho_2)
 f_1(x_1,\rho_1)f_2(x_2,\rho_2)\sigma_{2\to 2} =\nonumber \\
  \int d^2bd^2\rho_1d^2\rho_2f_1(x_1,\rho_1)f_2(x_2,\rho_2)\sigma_{2\to 2}=f_1(x_1)f_2(x_2)\sigma_{2\to 2}.& &
 \eea
 Here at the last step we used the relation between diagonal GPD and PDF:
 $\int d^2\rho f_j(x,\rho,Q^2)= f_j(x,Q^2)$.
 
 At the same time,  as soon as one wants to describe  the structure of the final state in production of say heavy particles, it is important to know whether a dijet production occurs at   different average impact parameters  than the
  minimum--bias interactions. We will argue below that 
 dijet trigger selects, in average,  a factor of two smaller impact parameters.
 This implies   that the multijet activity, energy flow  should be much stronger in these events than in the minimum--bias events.
 Obviously,  the magnitude of the enhancement does depend on the transverse distribution of partons and on correlation between the partons in the transverse plane.
 This information becomes available now.

 \section{Transverse structure of the nucleon wave function}
 The basis  for  the quantitative analysis of the transverse nucleon structure  is provided by  the QCD factorization theorem for exclusive vector meson (VM) production\cite{Collins:1996fb} which states that in the leading twist approximation  the differential cross section of the process  $\gamma^*_L + p \to VM + p$ is given  by the  convolution of the hard block, meson wave function and generalized gluon parton distribution, $g(x_1,x_2, t\left|\right. Q^2) $,
 where $x_1, x_2$  are the 
longitudinal momentum fractions of the emitted and absorbed gluon
  (we discuss here only the case of small x which is of relevance for the LHC kinematics).  Of particular interest 
is the generalized parton distribution (GPD) in the ``diagonal'' case, $g(x, t | Q^2)$, where $x_1=x_2$ and denoted by $x$, and the momentum transfer to the 
nucleon is in the transverse direction, with 
$t = -\bf{\Delta}_\perp^2$ (we follow the notation of
Refs.~\cite{Frankfurt:2003td,Frankfurt:2010ea}). This function reduces to the 
usual gluon density in the nucleon in the limit of zero momentum 
transfer, $g(x, t = 0| Q^2) = g(x| Q^2)$. Its two-dimensional
Fourier transform
\begin{equation}
g(x, \rho | Q^2) \; \equiv \; \int\!\frac{d^2 \Delta_\perp}{(2 \pi)^2}
\; e^{i (\bm{\Delta}_\perp {\bm\rho })}
\; g(x, t = -{\bm{\Delta}}_\perp^2 | Q^2)
\label{gpdrho_def}
\end{equation}
describes the one--body density of gluons with given $x$ in transverse space,
with $\rho \equiv |\bm{\rho}|$ measuring the distance from the 
transverse center--of--momentum of the nucleon, and is normalized
such that $\int d^2\rho \, g(x, \rho | Q^2) \;\; = \;\; g(x|Q^2). $
It is convenient to separate the information on the total 
density of gluons from their spatial distribution and parametrize
the GPD in the form
\be
g(x, t | Q^2) \;\; = \;\; g(x | Q^2) \; F_g(x, t | Q^2) ,
\ee
where the latter function satisfies $F_g(x, t =0| Q^2) = 1$ and is known as 
the two--gluon form factor of the nucleon. Its Fourier transform describes 
the normalized spatial distribution of gluons with given $x$,
\be
F_g (x, \rho | Q^2) \; \equiv \; \int\!\frac{d^2 \Delta_\perp}{(2 \pi)^2}
\; e^{i (\bm{\Delta}_\perp \bm{\rho})}
\; F_g (x, t = -{\bf{\Delta}}_\perp^2 | Q^2) ,
\label{rhoprof_def}
\ee
with $\int d^2\rho \, F_g (x, \rho | Q^2) = 1$ for any $x$.

The QCD factorization theorem predicts that the t-dependence of the VM production should be a universal function of $t$ for fixed $x$ 
(up to small DGLAP evolution effects).
Indeed the t-slope of the $J/\psi$ production is  practically $Q^2$ independent, 
while the t-slope of the production light vector mesons approaches that of $J/\psi$ for large $Q^2$.
The $t$--dependence of the measured differential cross sections of
exclusive processes at $|t| < 1 \, \mbox{GeV}^2$ is commonly
described either by an exponential, or by a dipole form inspired 
by analogy with the nucleon elastic form factors. Correspondingly,
we consider here two parametrizations of the two--gluon form factor:
\be
F_g (x, t|Q^2) \;\; = \;\; 
\left\{ \begin{array}{l}
\displaystyle
\exp (B_g t/2) ,
\\[2ex]
\displaystyle 
(1 - t/m_g^2)^{-2} ,
\end{array}
\right.
\label{twogl_exp_dip}
\ee
where the parameters $B_g$ and $m_g$ are functions of $x$ and $Q^2$.
The two parametrizations give very similar results if the functions 
are matched at $|t| = 0.5 \, \mbox{GeV}^2$, where they are best 
constrained by present data (see Fig.~3 of Ref.~\cite{Frankfurt:2006jp});
this corresponds to \cite{Frankfurt:2010ea}
\be
B_g \;\; = \;\; 3.24/m_g^2 .
\label{dip_exp}
\ee
The analysis of the HERA exclusive data leads to 
\bea
B_g (x) = B_{g0} \; + \; 2 \alpha'_g \; \ln (x_0/x) , 
\label{bg_param}
\eea
where $x_0 = 0.0012, 
B_{g0} = 4.1 \; ({}^{+0.3}_{-0.5}) \; \mbox{GeV}^{-2}, \alpha'_g = 0.140 \; ({}^{+0.08}_{-0.08}) \; \mbox{GeV}^{-2}$ for $Q_0^2\sim $ 3 GeV$^2$.  For fixed $x$, $B(x,Q^2)$ slowly decreases with increase of $Q^2$  due to the DGLAP evolution 
\cite{Frankfurt:2003td}. The uncertainties in parentheses represent a rough estimate
based on the range of values spanned by the H1 and ZEUS fits, 
with statistical and systematic uncertainties added linearly. This estimate does not include possible contributions to  $\alpha'_g$ due to the contribution of the  large size configurations in the vector mesons and changes in the evolution equation at 
$-t$ comparable to the intrinsic scale.  Correcting for these effects may lead to a reduction of $\alpha'_g$ and hence to a slower increase of 
the area occupied by gluons with decrease of $x$.

It is worth noting here that the popular Monte Carlo description of the pp collisions at the collider energies - PYTHIA uses x-independent transverse distribution of partons described by the sum of two exponentials. This distribution roughly  equivalent to the dipole parametrization with $m^2\approx  2\mbox{GeV}^2$ \cite{Siodmok} which is hardly consistent with the data on $J/\psi$ photoproduction, see dashed line in Fig. 3. For smaller $x$ the difference is even larger since the transverse size increases with decrease of $x$ -- see Eq. 7.
\begin{figure}[t]  
  \vspace*{-0.3cm}
   \hspace{2cm}\includegraphics[width=0.6\textwidth]{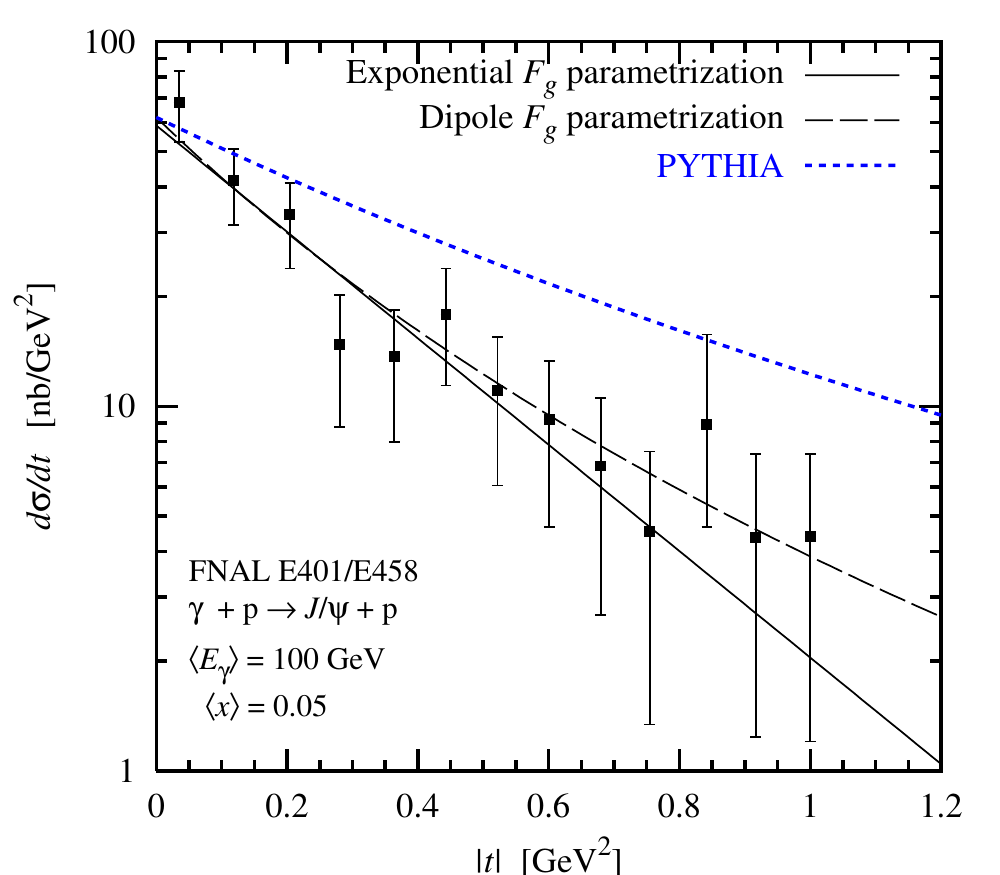} 
   \caption{ $t$-dependence of the exclusive $J/\psi$ photoproduction data
from the FNAL E401/458 experiment \cite{Binkley:1981kv}.
Solid line: $t$--dependence obtained with the exponential
parametrization of the two-gluon form factor, Eq.5
(the slope of the $J/\psi$ cross section is $B_{J/\psi} = B_g
+ \Delta B$, where $\Delta B \approx 0.3 \, \mbox{GeV}^{-2}$
accounts for the finite size of the $J/\psi$;
see \cite{Frankfurt:2010ea} for details).
Dashed line: $t$--dependence obtained with the dipole
parametrization, Eq.5. Dotted line: $t$--dependence
obtained with PYTHIA, effectively corresponding to a dipole
form factor with $m^2 \approx 2 \, \mbox{GeV}^2$.
}
       \label{psifig}
 \end{figure}

\section{Impact parameter distribution of proton--proton collisions with dijet production}
\label{sec:impact}
\begin{figure}[t]
\hspace{2cm}\includegraphics[width=.50\textwidth]{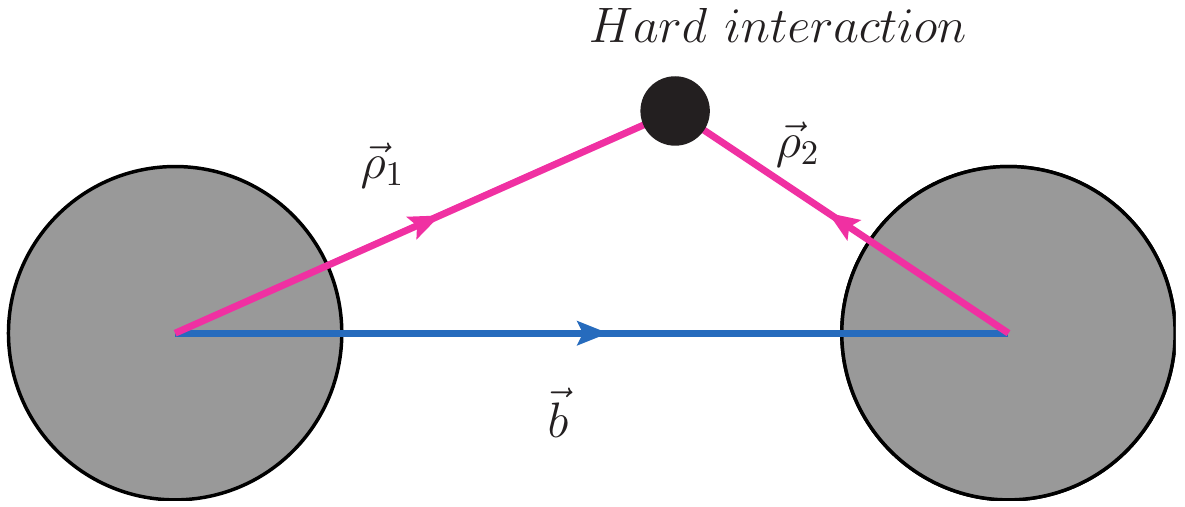}
\caption[]{Overlap integral of the transverse spatial parton
distributions, defining the impact parameter distribution of
$pp$ collisions with a hard parton--parton process, Eq.~(\ref{P_2_def}).}
\label{fig:overlap}
\end{figure}

Using the information on the transverse spatial distribution 
of partons in the nucleon, one can infer the distribution of
impact parameters in $pp$ collisions with hard parton--parton
processes \cite{Frankfurt:2003td}.  It is given by the overlap of two parton wave function as depicted in Fig. \ref{fig:overlap}.

The probability distribution of 
$pp$ impact parameters in events with a given hard process, $P_2 (x_1, x_2, b|Q^2)$,   is given by the ratio of the cross section at given $b$ and the cross section integrated over $b$.
As a result 
\bea
P_2 (x_1, x_2, b|Q^2) &\equiv&
\int \! d^2\rho_1 \int \! d^2\rho_2 \; 
\delta^{(2)} (\bm{b} - \bm{\rho}_1 + \bm{\rho}_2 )
\nonumber \\
&&\times F_g (x_1, \rho_1 |Q^2 ) \; F_g (x_2, \rho_2 |Q^2) \, ,
\label{P_2_def}
\eea
which obviously satisfies the  normalization condition
\be
\int d^2b \, P_2 (x_1, x_2, b |Q^2) \;\; = \;\; 1.
\ee
This distribution represents an essential tool for  phenomenological 
studies of the underlying event in $pp$ 
collisions \cite{Frankfurt:2003td,Frankfurt:2010ea}.
We note that the concept of impact parameter distribution is also used 
in MC generators of $pp$ events with hard 
processes \cite{PYTHIA,HERWIG}, albeit without making 
the connection with GPDs, which allows one to import information on 
transverse nucleon structure obtained in the independent measurements.

For the two  parametrizations of Eq.~(\ref{twogl_exp_dip}),  
 Eq.~(\ref{P_2_def}) leads to  ( for $x\equiv x_1 = x_2$) 
 \be
P_2 (x, b| Q^2) \; = \; \left\{
\begin{array}{l}
\displaystyle
(4\pi B_g)^{-1} \, \exp [-b^2/(4 B_g)] ,
\\[2ex]
\displaystyle
[m_g^2 /(12\pi)] \, (m_g b/2)^3 \, K_3 (m_{g} b) ,
\label{P_2_exp_dip}
\end{array}
\right.
\label{p2}
\ee 
where the parameters $B_g$ and $m_g$ are taken at the appropriate
values of $x$ and $Q^2$.

The derived distribution should be compared to the distribution of the minimum--bias inelastic collisions which could be expressed through  $\Gamma (s, b)$ that  is the profile function of the $pp$ elastic
amplitude ($\Gamma (s, b)=1$ if the interaction is completely absorptive at given $b$)
\be
P_{\mbox{in}} (s, b) \;\; = \;\; 
\left[ 1 - |1 - \Gamma (s, b)|^2\right]\,  / \sigma_{\mbox{in}}(s) ,
\label{P_in_def}
\ee
where 
$\int d^2 b \, P_{\mbox{in}} (s, b) = 1$.

Our numerical studies indicate that  the
impact parameter distributions with the jet trigger (Eq.\ref{P_2_exp_dip})  are much
narrower than that in minimum--bias inelastic events at the same energy (Eq.\ref{P_in_def})
and that  $b$-distribution for events with a dijet trigger is a very weak function of the $p_T$ of the jets or their rapidities, see Fig.5.  For example for the case of the $pp$ collisions at $\sqrt{s}= \mbox{7 GeV} $ we find the  median value of $b$,
$ b_{median} \approx$ 1.18 fm  and $ b_{median} \approx$ 
0.65 fm for minimum--bias and dijet trigger events (Fig. 5b)\cite{Frankfurt:2010ea}. Since the large impact parameters give the dominant contribution to $\sigma_{inel} $ our analysis indicates that
there are two pretty distinctive classes of pp collisions  - large $b $ collisions which are predominantly soft and and central collisions with strongly enhanced rate of hard collisions. We refer to this pattern as the two transverse scale picture of $pp$ collisions at collider energies  \cite{Frankfurt:2003td}.

A word of caution is necessary here. The transverse distance $b$ for dijet events is defined as the distance between the transverse centers of mass of two nucleons. It may not coincide with  $b$ defined for soft interactions where soft partons play an important role. For example, if we consider dijet production due to the interaction of two partons with $x\sim 1$, $\rho_1,\rho_2 \sim 0$ since the transverse center of mass coincides with transverse position of the leading quark in the $x\to 1$ limit.
  As a result $b$ for the hard collision will be close to zero. On the other hand the rest of the partons may interact in this case at different transverse coordinates. As a result,  such configurations may    contribute to the inelastic $pp$  cross section at much larger  $b$ for the soft interactions. However for the parton collisions at  $x_1,x_2 \ll 1$   
the recoil effects  are small and so two values of $b$ should be close.

  \begin{figure}
\begin{tabular}{ll}
\includegraphics[width=.45\textwidth]{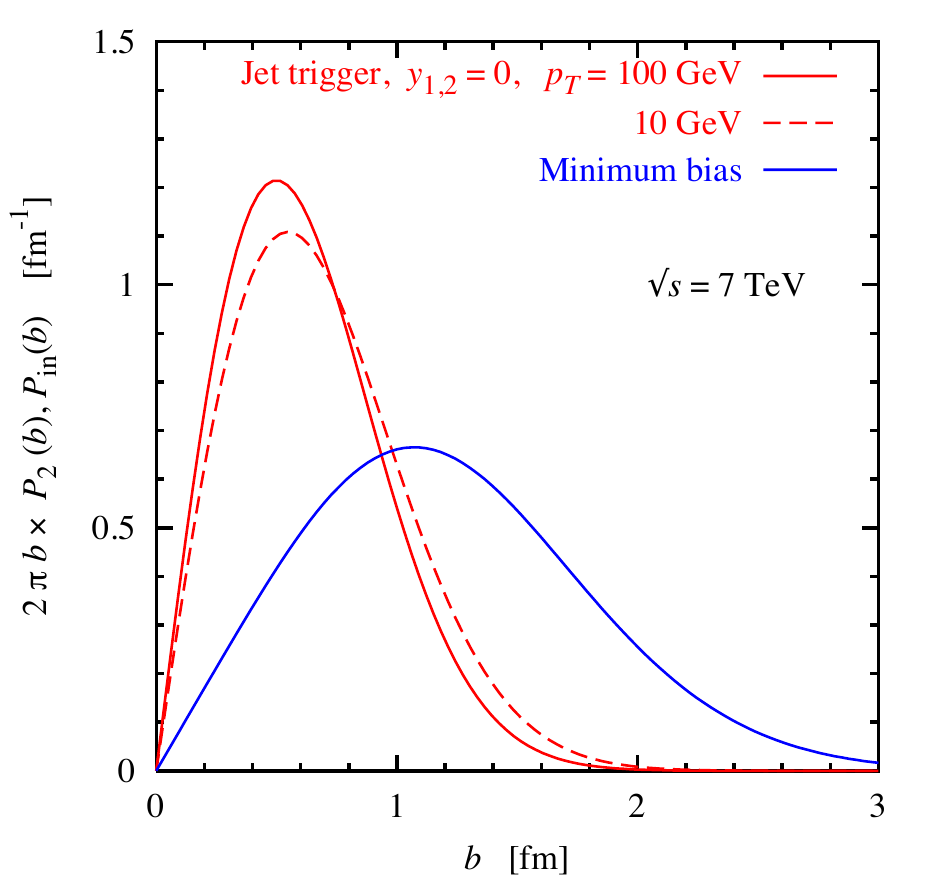} &
\includegraphics[width=.45\textwidth]{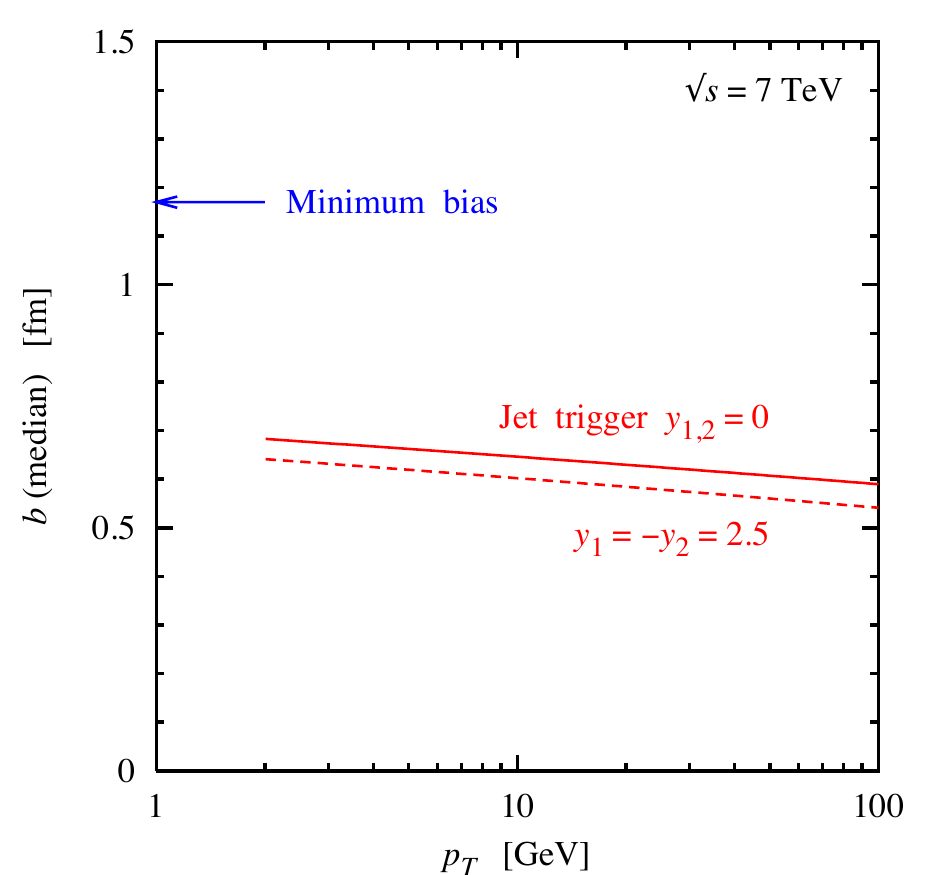}
\end{tabular}
\caption[]{(a) Impact parameter distributions of inelastic $pp$ 
collisions at $\sqrt{s} = 7 \, \mbox{TeV}$.
\textit{Solid (dashed) line:} Distribution of events with a 
dijet trigger at zero rapidity, $y_{1, 2} = 0$, for $p_T = 100 \, (10) \, \mbox{GeV}$ cf. Eq.~(\ref{P_2_exp_dip}). 
\textit{Dotted line:} Distribution of minimum--bias inelastic events,cf. \ Eq.~(\ref{P_in_def}). (b) Dependence of median $b$ on $p_T$ for different rapidities of the dijets.}
\label{sob}
\end{figure}

The present $pp$ LHC data already provide important tests of this picture. Let us consider production of the hadron (minijet) with momentum $p_T$. The observable of interest
here is the transverse multiplicity, defined as the multiplicity
of particles with transverse momenta in a certain angular region 
perpendicular to the transverse momentum of the trigger particle or jet
(the standard choice is the interval $60^\circ < |\Delta \phi| < 120^\circ$ 
relative to the jet axis; see Ref.~\cite{Affolder:2001xt} for an
illustration and discussion of the experimental definition).  
In the central collisions one expects a much larger transverse multiplicity due to the presence of multiple hard and soft interactions. 
At the same time 
the enhancement 
should be a weak function of $p_T$ in the region where main contribution is given by the hard mechanism \cite{Frankfurt:2003td,Frankfurt:2010ea}. The predicted increase and eventual flattening of the 
transverse multiplicity agrees well with the pattern observed in the
existing data. At $\surd s = 0.9\, \textrm{TeV}$ the transition occurs 
approximately at 
$p_{T, {\rm crit}} \approx 4\, \textrm{GeV}$,
at $\surd s = 1.8\, \textrm{TeV}$ at 
$p_{T, {\rm crit}} \approx 5\, \textrm{GeV}$,
and  at $p_{T, {\rm crit}} = 6-8\, \textrm{GeV}$  for $7\, \textrm{TeV}$ \cite{Chatrchyan:2011id,Aad:2011qe}. We thus conclude that the minimum $p_T$ 
at which  particle production due to hard collisions starts to dominate  significantly  increases with the collision energy.
This effect is likely to be related the onset of the high gluon density regime in the central $pp$ interactions since with an increase of incident energy partons in the central $pp$ collisions propagate through stronger and stronger gluon fields. 

Many further tests of the discussed picture which are suggested in Ref.~\cite{Frankfurt:2010ea} will be feasible in a near future. They include
(i) Check that the transverse multiplicity  does not depend
 on rapidities of the jets, (ii) Study of the  multiplicity  at 
$y < 0$ for events with   jets at $y_1 \sim  y_2 \sim 2$. This would allow to check that the     transverse  multiplicity is universal  and that multiplicity in the   away and the  towards regions is similar to the  transverse multiplicity for $ y \le 0$. (iii) Studying whether transverse multiplicity is the same for quark and  gluon induced jets. Since the gluon radiation for production of  $ W^{\pm},Z$ is smaller than for the gluon dijets, a  subtraction of the radiation effect mentioned below is very important for such a comparisons. 

Note that the contribution of the jet the fragmentation to  the transverse cone as defined in the experimental analyses  is small but not negligible especially at smaller energies
($\sqrt{s}=0.9 \mbox{TeV}$). It would be desirable to use a more narrow transverse cone, or subtract the contribution of the jets fragmentation. Indeed,   the color flow  contribution \cite{Dokshitzer:1991wu} leads to a small residual increase of the transverse multiplicity  with $p_T$.
However the jet fragmentation  effect depends on $p_T$ rather than on $\sqrt{s}$. Hence it does not contribute to the growth of the transverse multiplicity, which is   a factor of $\sim 2$ between $\sqrt{s}=0.9 \mbox{TeV}$ and $\sqrt{s}=7.0 \,\mbox{TeV}$.  In fact, a subtraction of the  jet fragmentation contribution would somewhat increase the rate of the increase of the   transverse  multiplicity in the discussed energy interval. This allows to obtain the lower limit for the rate of the increase of the multiplicity in the central ($<b>\sim$ 0.6 fm) $pp$ collisions of $s^{0.17}$. It is    a bit  faster than the $s$ dependence of multiplicity in the central heavy ion collisions.

\begin{figure}[t]
\hspace{2cm}\includegraphics[width=.60\textwidth]{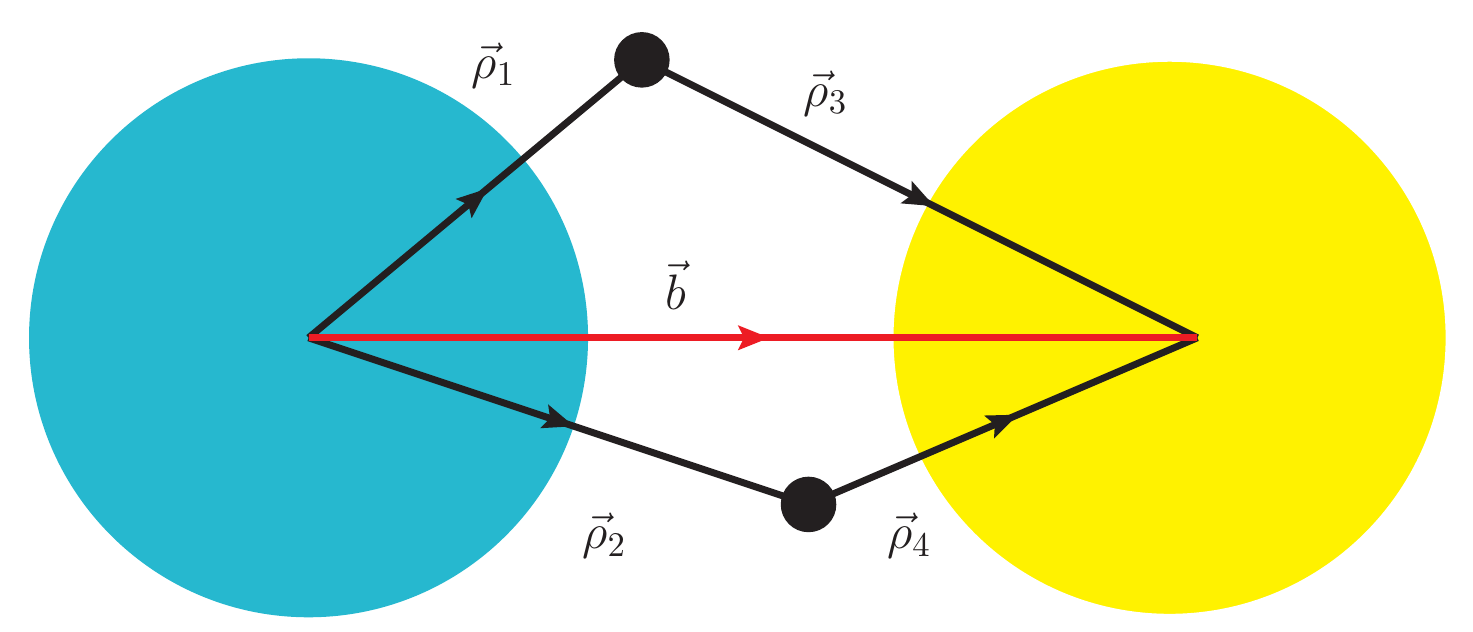}
\caption[]{Geometry of two hard collisions in impact parameter picture. }
\label{fig:overlap4}
\end{figure}

\section{Multiparton interactions}
Probability of  the multiparton interactions which result in two hard collisions  grows rapidly with the increase of the incident energy.  Understanding of such processes is important  for  detailed understanding of the high energy QCD dynamics
as well as for 
practical purposes  --  estimating backgrounds for the searches for new particles. 
By exploring the difference in scales between soft and hard QCD processes
 and space-time structure of Feynman diagrams 
 we derive within pQCD the general formulae for the two dijet production
in pp collisions \cite{Blok:2010ge,Blok:2011bu} and find that it contains two contributions. The contribution which dominates in a wide range of $x_i$ is  the $4\to 4$ process  which matches the intuitive geometric picture depicted in  Fig. \ref{fig:overlap4}.
However, a consistent pQCD treatment requires that one also takes into account  $3\to4$ double hard interaction processes that occur as an interplay between large- and short-distance parton correlations\cite{Blok:2011bu}. Such contributions  are not taken into consideration by approaches inspired by the parton model picture. This contribution takes into account correlations between the partons induced by the pQCD evolution. Our analysis indicates that this contribution becomes important only for $x\le 10^{-3}$.

 The $4
\to 4$  cross section for the collisions of hadrons "a" and
"b" has the form\cite{Blok:2010ge}:
\begin{eqnarray}\label{eq:main_form}
d\sigma_4 &=& \int \frac{d^2\overrightarrow{\Delta}}{(2\pi)^2}\int dx_1\int  dx_2\int dx_3\int  dx_4  \nonumber\\[10pt]
&\times &D_a(x_1,x_2,p_1^2,p_2^2, \overrightarrow{\Delta}) D_b(x_3,x_4,p_1^2,p_2^2,-\overrightarrow{\Delta})
\displaystyle{\frac{d\sigma^{13}}{d\hat t_1} \frac{d\sigma^{24}}{d\hat t_2}} d\hat t_1d\hat t_2.    
\label{b1}
\end{eqnarray}
Here $D_\alpha(x_1,x_2,p_1^2,p_2^2,  \overrightarrow{\Delta})$
are the new  "double" GPDs
 for hadrons "a" and
"b":
\begin{eqnarray}
&&\hspace{-1cm} D(x_1,x_2,p^2_1,p^2_2,\overrightarrow{\Delta})
=\sum_{n=3}^{\infty}\int
\frac{d^2k_1}{(2\pi)^2}\frac{d^2k_2}{(2\pi)^2}\theta
(p_1^2-k_1^2)\theta (p_2^2-k_2^2)
\nonumber\\[10pt] &&
\times \int \prod_{i\ne
1,2}\frac{d^2k_i}{(2\pi)^2}\int^1_0\prod_{i\ne
1,2} dx_i\, (2\pi)^3\delta( \sum_{i=1}^{i=n} x_i-1)
\delta (\sum_{i=1}^{i=n} \vec k_i)
\nonumber\\[10pt]
&&
\times 
\psi_n (x_1,\vec k_1,x_2,\vec k_2,.,\vec k_i,x_i..)\psi_n^+(x_1,\overrightarrow{k_1}+\overrightarrow{\Delta},x_2,\overrightarrow{k_2}
-\overrightarrow{\Delta},x_3, \vec k_3,...) 
.
\label{b2}
\end{eqnarray}
Note that this distribution is diagonal in the space of all
partons except the two partons involved
 in the collision. Here $\psi$ is the parton
wave function normalized to one in the  usual way. An appropriate
summation over color and Lorentz indices is implied.
\par
Within the parton model approximation the cross section has the form:
 \be
 \sigma_4= \sigma_1\sigma_2/\pi R^2_{\rm int},
 \label{b3}
 \ee
where $\sigma_1$ and $\sigma_2$ are the cross sections of two independent hard binary parton interactions. The factor $\pi R^2_{\rm int}$ characterizes the transverse area occupied by the
partons participating in the two hard collisions. It also includes effect  of possible longitudinal correlations between the partons. 

Eq.~(\ref{b1}) leads to  the general model
independent expression for 
 \be
\frac{1}{\pi R^2_{\rm int}}=\int
 \frac{d^2\overrightarrow{\Delta}}{(2\pi)^2}D(x_1,x_2,-\overrightarrow{\Delta})
 D(x_1,x_2,\overrightarrow{\Delta}),
 \label{b33}
 \ee
in terms of two-parton GPDs. 

In the independent particle approximation which is  used in all Monte Carlo models with  multiparton interactions, the  two-parton GPD is equal to the product of single particle GPDs discussed in section 2. Using parametrization of 
Eq.~(\ref{P_2_exp_dip}) one finds \cite{Frankfurt:2003td,Blok:2010ge}
\be \frac{1}{\pi R^2_{\rm int}}=\int
\frac{d^2\Delta}{(2\pi)^2}F_{g}^4(\Delta)=\frac{m^2_g}{28\pi},
\label{b4}
\ee
 which leads to approximately a factor of two smaller cross section than the one observed at the Tevatron:
 $\pi R^2_{\rm int}\approx 34 mb$  as compared to the experimental value of
  $\pi R_{int}^2 \approx 15 mb $.  The $3\to 4$ processes  plays a minor role in the Tevatron kinematics and do not allow to solve this discrepancy. 
  
  A fix  implemented in the PYTHIA is to use a much more narrow distribution in $\rho$  - effectively a dipole with $ m^2= 2 \mbox{GeV}^2$. This does decrease $\pi R_{int}^2$ to the value observed at  the Tevatron.  However it is a factor of two smaller than the one  determined from  the analyses
   of the hard exclusive processes  -- see discussion in section 2 and  in particular Fig.3.  Correspondingly in this model the difference between median $\rho^2$ for minimum--bias processes   and processes with dijet trigger is a factor of 2 larger than what follows from the analysis of the HERA data, cf. Fig. \ref{sob}.

 In principle there is a question of 
  how well  the separation of the $2\to 4$ processes  was performed in the experimental studies.
  However the most recent D0 analysis\cite{Abazov:2011rd}  seems to indicate that the $2\to 4$ contribution in the kinematics used to determine $\pi R_{int}^2$ is very small.
  
 This appears to leave us with only one possibility - presence of significant parton - parton correlations at a nonperturbative scale.  Currently we are performing estimates of these correlations. We find  that the correlations are indeed  large and may explained the enhancement   we discussed in this section.

 \section{Fluctuations of the gluon field and high multiplicity  events at LHC}
 Strength of the gluon field should depend on the size of the quark configurations.
 For example,  the gluon field in the  small configurations should be  strongly screened -- the  gluon density much smaller than average. It is possible to extract from the comparison of the diffractive processes: $\gamma^\ast_L + p \rightarrow V + X$ and $\gamma^\ast_L + p \rightarrow V + p$  the dispersion  of the gluon strength at small x \cite{Frankfurt:2008vi}:
 \begin{eqnarray}
\omega_g \;\; \equiv \;\; 
\frac{\langle G^2 \rangle - \langle G \rangle^2}{\langle G \rangle^2}
\;\; = \;\; 
\left. \frac{d\sigma_{\gamma^*+p\to VM +X}}{dt} \right/ \left.
\frac{d\sigma_{{\gamma^*+p\to VM +p}}}{dt} \right|_{t=0} .
\label{fluct}
\end{eqnarray}
The HERA data indicate that for $Q^2\sim 3 \mbox{GeV}^2$ and $x\sim 10^{-3}$
$\omega_g  \sim 0.15 \div 0.2$ which is rather close to the value for the analogous ratio for the soft diffraction which measures fluctuations of overall strength of the {\it soft} hadronic interactions.

How can one   probe the gluon fluctuations in $pp$ collisions? Let us  consider multiplicity of an inclusive  hard process -- dijet,...  as a function of  some cuts -- for example overall hadron multiplicity: M (trigger) and build the ratio 
\begin{equation}
R={M(trigger)\over M(minimum--bias)}.
\end{equation}
 If there are no fluctuations of the parton densities, the  maximal  value of $R$
  is reached if the trigger selects collisions at small impact parameters $b\sim 0$.
  Using Eq. \ref{p2} we find \cite{Strikman:2011ar}
  \begin{equation}
  R= P_2(0)\sigma_{in}(pp)= {m_g^2\over 12 \pi}\sigma_{in}(pp)\approx 4.5.
   \end{equation}
   Any larger enhancement of $R$ could arise only from the fluctuations of the gluon density per unit area.
   
   The first measurement which maybe relevant for addressing the question of fluctuations was reported by ALICE \cite{Kramer}. The multiplicity of $J/\psi$ was studied as a function of the  multiplicity in the central detector, namely  
   \begin{equation}
   dN^R_{ch}/d\eta= {dN_{ch}/d\eta_{\eta=0}\over \left<dN_{ch}/d\eta_{\eta=0}\right>},
   \end{equation}
   for $dN^R_{ch}/d\eta \le 5$.
   It was found that  $R$ increases with increase of $dN^R_{ch}/d\eta$ reaching values $\approx 5$ for $dN^R_{ch}/d\eta \sim 4$. 
   This number is close to what we estimated above. Any further increase of $R$ would require presence of the fluctuations in transverse gluon density. An enhancement above the $b=0$ effect is given by the factor
   \be
   R_{fl}= {g_{N}(x_1,Q^2|n)g_{N}(x_2,Q^2|n)\over g_{N}(x_1,Q^2)g_{N}(x_2,Q^2)}
 { \left< S\right>\over S}.
\ee
Here $n$ labels configurations selected by the trigger, and  $S$ is the area of the transverse overlap. In principle $R_{fl}$ could reach very  large values. For example, if we consider a collision of two protons in  cigar shape configurations with the same gluon density for different orientations of the protons,  the enhancement would be proportional to  the ratio of the principal axes of the ellipsoid.
Another mechanism for the enhancement of $R_{fl}$ is presence of the dispersion in the gluon density with $\omega_g \sim 0.15 \div 0.2$ (Eq. \ref{fluct}) which leads to  a few percent probability for the gluon field to be a factor 1.5 larger than average.

These observations maybe of relevance for the discussion of the high multiplicity (HM) events studied by the CMS \cite{Khachatryan:2010gv}.  In the analysis  very rare events were selected which have the overall multiplicity for $|\eta| < 2.4$ of at least  a factor of  $\ge 7$  larger than  the minimum--bias events. Probability of such events is very small:
\be
P_{HM}\approx 10^{-5} \div 10^{-6}.
\ee
  The two-particle correlations were measured as a function of the distance in the pseudorapidity - $\Delta \eta$ and the azimuthal angle   - $\Delta \phi$.
   Three types of correlations were observed : (a) very strong local correlation for $\Delta \eta\sim 0, \Delta \phi \sim 0$, (b) strong correlation for $\Delta \phi \sim \pi$
   for a wide range of $\Delta \eta$, (c) a weak correlation for $2 < |\Delta\eta | < 4.8,  \Delta \phi \sim 0$ -- so called ridge.
   
   The first question to address is how to get such a large multiplicity. It is pretty obvious that such events  should originate from   very central collisions. Based on our knowledge of $P_2(b)$ we  find that the probability of the collisions at $b < 0.2 fm$ is $\sim $2\%. Using information about dispersion of fluctuations of the gluon fields we estimate the probability of fluctuation where  both nucleons  have $g > 1.5 g_N(x)$ is $\ge 10^{-3}$.  So a natural guess is that the CMS trigger selected central collisions with enhanced gluon fields in both nucleons. This should lead to a much higher rate of jet production  per event.
    
    Indeed our analysis of the HM data  indicates  presence of a large total excess transverse momentum in the $\Delta \phi \sim \pi $   region: $p_t^{balance} \ge \,   15 \mbox{GeV/c}$. Presumably it is due to production of two back to back jets with the trigger jet generating the narrow same side correlation. Qualitatively, a large probability of the dijets maybe due to the combination of centrality and the gluon density fluctuation. 
    
   Note also that the increase of the multiplicity due to selection of $b\sim 0$ and 
   selection of $b\sim 0$  and enhanced dijet production is not sufficient to generate 
        a factor of 7 increase in the multiplicity - without of the gluon density fluctuations these two effects typically lead to $N_{ch} \sim 70$.   The $g >  g_N(x)$ gluon  fluctuations would naturally lead to a  further increase of $N_{ch}$.
    
    The same side ridge could originate  from the the string effect \cite{Dokshitzer:1991wu}. This could be tested by studying  collisions with production of dijets with $p_T\sim   \mbox{15 GeV/c}$ without HM trigger. Alternative  mechanism would be  fluctuations of  the transverse shape of the colliding nucleons plus presence of the absorptive effects for $p_t \le 3 
    \mbox{GeV/c}$. Such a scenario   appears  quite natural for the high density mechanism we discuss here.

    \section{Where does the non-linear regime set in?}
 
 In the leading log approximation one can derive a relation between the QCD evolution equations and the target  rest frame picture of the interaction of 
 small  color dipoles with targets expressing it through the gluon density in the target, see  \cite{Frankfurt:1996ri}
and refs. therein. Matching the behavior of the dipole cross section in the pQCD regime and of large size dipoles in the regime of soft interaction,  it is possible to write interpolation formulae for the dipole - nucleon cross section for all dipole sizes and describe
 the total cross section of DIS at HERA. 
 
 To determine how close is the interaction strength to the maximal allowed by the unitarity it is necessary to consider the amplitude of the $q\bar q$ dipole - nucleon interaction in the impact parameter space:
 \begin{equation}
 \Gamma_{q\bar q}(s,b)={1\over 2is}{1\over (2\pi)^2} \int d^2\vec{q}e^{i\vec{q}\vec{b}}A_{q\bar q - N}(s,t),
 \end{equation}
 where $ A_{q\bar qN}(s,t)$ is the elastic amplitude of the $q\bar q$ dipole - nucleon scattering normalized to $\it{Im} A_{q\bar qN}= is_{q\bar q - N}  \sigma_{tot}(q\bar q -N)$. The limit $ \Gamma_{q\bar q}(s,b)=1$ corresponds to the regime of the complete absorption - black disk regime (BDR) - the maximal strength allowed by the S-channel unitarity.
  
 The   $t$ dependence of the $ q\bar q$ dipole - nucleon elastic scattering amplitude can be obtained from the studies of the exclusive vector meson production in the regime where QCD factorization theorem for the exclusive processes allows to  relate the  $t$ dependence of the amplitude to the $t$-dependence of the gluon GPDs. 

\begin{figure}[h]  
   \centering
   \includegraphics[width=0.9\textwidth]{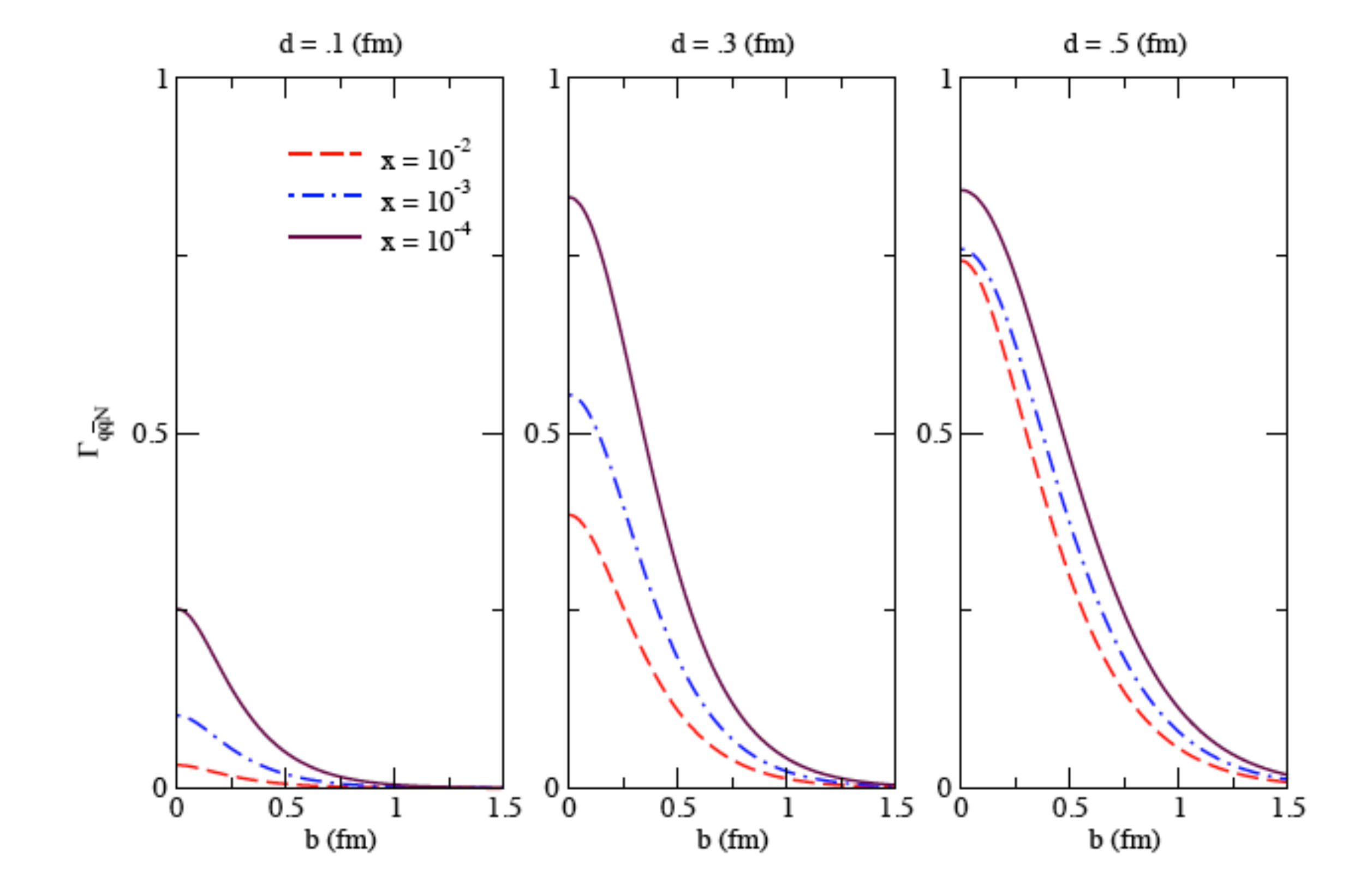} 
 \caption{Impact parameter distribution of $q\bar q $ dipole interaction with protons adopted from \cite{Rogers:2003vi}.}
    \label{impactfig}
 \end{figure}

Combining this information with the information on the total cross section of the dipole - nucleon interaction allows us to determine  $ \Gamma_{q\bar q _N}(s,b)$  as function of the dipole size.  A sample of the results for $q\bar q$ dipole -proton  interaction which represent an update of the analysis of \cite{Rogers:2003vi}
is presented in Fig.\ref{impactfig}. For the case of color octet dipole $\Gamma_{inel}^{gg} = (9/4) \Gamma_{inel}^{q\bar q} $ leading to $\Gamma^{gg}$ much closer to one.  As a result the gluon induced  interactions are close to the BDR for a much larger range of the dipole sizes (this is consistent with the observation at HERA of a much larger probability of the diffraction in the gluon induce small x DIS processes).

Note also that $\Gamma =1/2 $ already corresponds to a probability of inelastic interaction of 3/4 which is close to one. One can also demonstrate  that the inelastic interactions get much larger corrections for the structure of  the final states than the total cross section, see discussion in \cite{Frankfurt:2011cs}.

In the case of the nuclear target the gain in the value of $\Gamma(b\sim 0)$
is rather modest due to the leading twist shadowing. The main gain in the nucleus case is due to a weak dependence of $\Gamma(b)$ on $b$
for a broad range of $b$.

\begin{figure}[h]  
   \centering
   \includegraphics[width=0.9\textwidth]{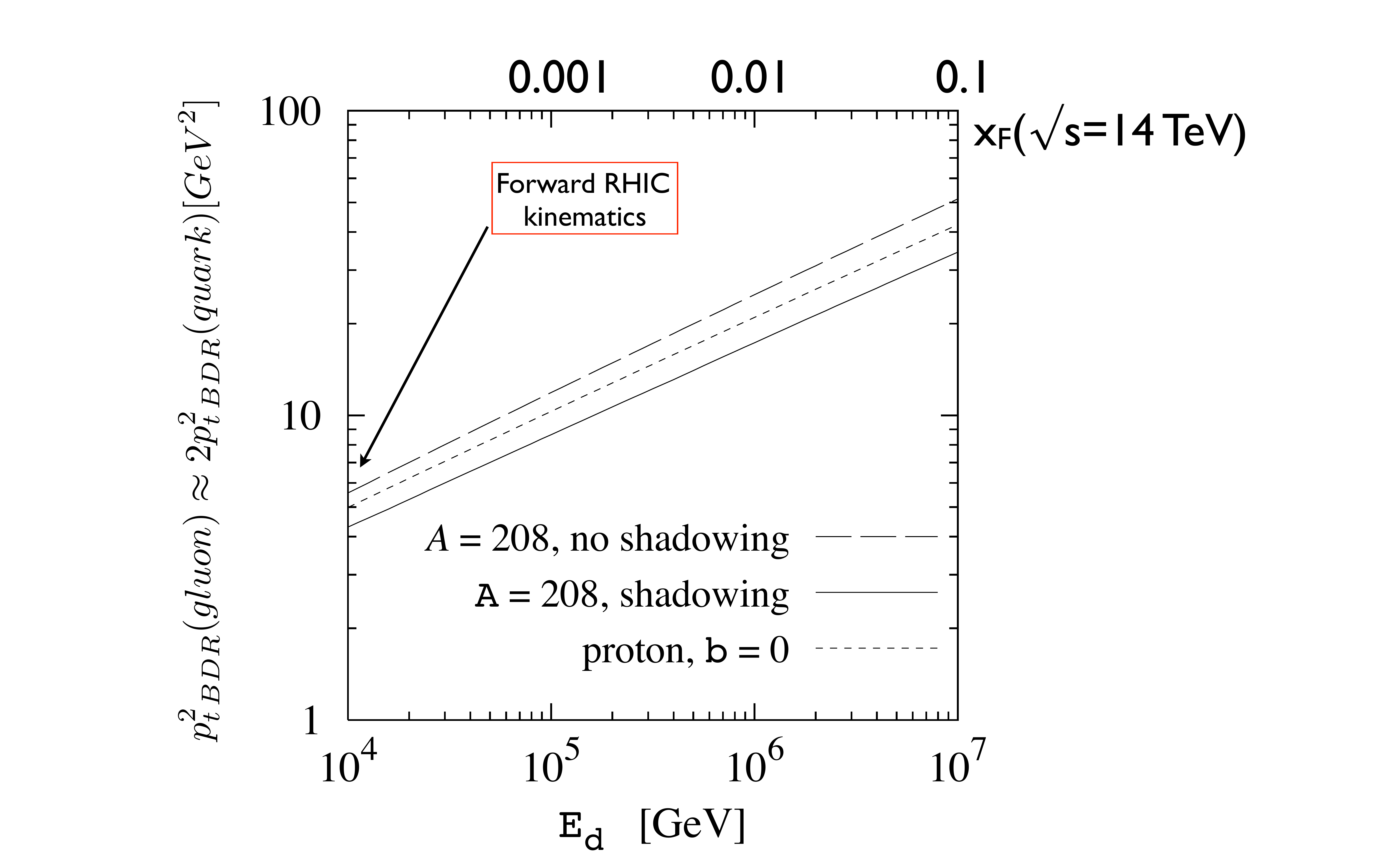} 
 \caption{The $p_T$ range where interaction is close to the BDR for the interaction of $q\bar q $ and color octet dipoles plotted as a function of the energy of the dipole and of $x$ of the interacting  parton for $pp$ interactions at $\sqrt{s}=14 \mbox{TeV}$}.
    \label{bdrpt}
 \end{figure}

Information about  $\Gamma_{q\bar q - N} (b)$ can be used to estimate the  range 
of the  transverse momenta for which probability of the inelastic interaction of parton is close to one \cite{Frankfurt:2003td}. The results of this analysis  for $b=0$ are presented in Fig.\ref{bdrpt}. One can see from the figure that interaction of gluons is close to the 
 BDR for  a wide range of virtualities for the central $pp$ collisions at the LHC.   This is because a parton in the nucleon with a given $x_1$   resolves the gluons in the second nucleon with $x_2$ down to $4 p_T^2/ x_{1}s$. For example taking   $x\sim 10^{-2} $,  $\sqrt{s}$=14 TeV  and  $\sim p_T^2 = 4 \mbox{GeV}^2$  we find    $x_{2}(min)=  10^{-4}$. 
 Though there are substantial uncertainties in this analysis  due to the use of the leading log DGLAP approximation and extrapolation of the gluon densities to very small x, the analysis provides   a reasonable estimate of the magnitude of $p_T(BDR)$. Note also that the range we find for the RHIC kinematics for the central $NA$ collisions is consistent with the effect of the suppression of the forward  pion production due to the onset of the BDR which we discuss in Sects. 8, 9.

 Consequently the LT approximation should be broken for a wide range of $x_1,x_2$ for gluon -- gluon interactions (these are $x$'s for which DGLAP works well for the  DIS at HERA).  A breakdown of the LT approximation maybe  of relevance for interpretation of the empirical observation made in a number of the Monte Carlo studies that to avoid contradictions with the data one needs to introduce a cutoff for  minimal $p_T $ of the  hard interactions. The cutoff  is  $p_T(min) \ge$ 3 GeV/c for $\sqrt{s}= \mbox{ 7 TeV}$ and grows with $s$.  In the Monte Carlo models the value of this cutoff is effectively driven by the requirement that the multiplicity of hadrons due to hard interactions remains below the total multiplicity observed experimentally. One can reach  similar results (avoiding questions about sensitivity to the hadronization
 mechanism) based on the calculation of the probability of hard interactions as a function of the impact parameter and requiring that it does not exceed $\Gamma_{inel}(b)$ which is known from the S-channel unitarity 
\cite{Rogers:2008ua,Rogers:2009ke}.  Note however that the MC models with $p_T(min)$ cutoff strongly reduce the interactions of the large $x$ partons with the target which contradicts the proximity to the BDR. Hence one may expect that such Monte Carlo models overestimate the cross section  of production of large $x_F$   hadrons, especially for the central collisions.

\section{Postselection effect in BDR - effective fractional  energy losses}
 It was demonstrate  in \cite{Frankfurt:2001nt} that in the BDR  interactions with the target select configurations in the projectile wave function where  the projectile's energy is split between constituents much more efficiently  than in the DGLAP regime. The simplest   example is inclusive production of the leading hadrons in DIS for $Q\le 2 p_T(BDR)$. Interactions with the target are not suppressed up to   $p_T\sim p_T(BDR)$, leading to selection of configurations in $\gamma^*$ where longitudinal fractions carried by quark and antiquark are comparable. The photon energy splits between the partons {\bf before} the collision. It is the interaction that  selects at different energies different set of configurations which are resolved. Hence we refer to this phenomenon as the  postselection. As a result to a first approximation the leading hadrons  are produced in the independent fragmentation of $q$ and $\bar q$:
\begin{eqnarray}
\bar{D}^{\gamma^{\ast}_{T}\to h}(z)=2 \int^{1}_{z}dy D^{h}_{q}(z/y) \frac{3}{4}(1+(2y-1)^2),
\end{eqnarray}
leading to a strong suppression of the  hadron production at $x_F> 0.3$.

In the case of a parton  of a hadron  projectile propagating through the nucleus 
near BDR effective energy losses were estimated in Ref.\cite{Frankfurt:2007rn}. For quarks they  are expected to be  of the order of 10  \% in the regime of the onset of BDR and larger deep inside this regime. Also the effective energy losses  are somewhat larger for gluons as the $g\to$ gg splitting is more symmetric in the light cone fractions than the
 qg splitting.

\section{Leading hadron production in hadron - nucleus scattering}
\label{single}
Production of leading hadrons with $p_t\sim \mbox{few \, GeV/c}$ in hadron - nucleus scattering at high energies provides a sensitive test of the  onset of the BDR dynamics. Indeed in this limit pQCD provides a good description the forward single inclusive pion production in $pp$  scattering at RHIC \cite{Werner}. At the same time it was found to overestimate grossly   the cross section of the pion production in $dAu$ collisions at RHIC in the same kinematics. The analysis of ~\cite{GSV} has demonstrated that the dominant mechanism of the single pion production in the $NN$ collisions in the kinematics  studied at RHIC is scattering of leading quarks of the nucleon  off the gluons of the target
with the median value of $x_g$ for the gluons to be in the range $x_g \sim 0.01 \div 0.03$ depending on the rapidity of the pion.  The nuclear gluon density for such $x$ is known to be close to the incoherent sum of the gluon fields of the individual nucleons since the coherent length in the interaction is rather modest for such $x$.  As a result the leading twist  nuclear shadowing effects can explain only a very small fraction of  the observed suppression \cite{GSV} and one needs a novel dynamical  mechanism to suppress generation of pions in such collisions. It was pointed out in \cite{GSV} that the energy fractional 
energy losses on the scale of 10\% give a correct magnitude of suppression of the inclusive spectrum due to a steep fall of the cross section with $x_F$ which is consistent with the estimates within the 
 postselection mechanism.
 
 An important additional information comes from the 
 measurement of the correlation of the leading pion production with production of the pion production at the central rapidities 
 \cite{star,phenix2}. This corresponds to the kinematics which receives   the dominant contribution from the scattering off gluons with $x_g\sim 0.01 \div 0.02$. The rate of the correlations for $pp$ scattering is consistent with pQCD expectations.
An extensive analysis performed in \cite{Frankfurt:2007rn} has demonstrated that the strengths of "hard forward pion"  -- "hard $\eta \sim 0$ pion"  correlations in 
$ dAu$ and in $pp$ scattering are similar. A  rather small difference in the pedestal originates from the multiple soft collisions. Smallness of the increase of the soft pedestal as compared to $pp$ collisions unambiguously  demonstrates that the dominant source of the leading pions is the $dAu$  scattering at large impact parameters. This conclusion is supported by the experimental observation \cite{Rakness} that the associated multiplicity of soft hadrons in events with forward pion is a factor of two smaller than in the minimum--bias $dAu$ events. A factor of two  reduction 
 factor is consistent with the estimate of  \cite{Frankfurt:2007rn} based on the analysis of the soft component of $\eta =0$ production for the forward pion trigger.
 Overall these data indicate that 
 (i) the dominant source of the forward pion production 
 is the $2\to 2$ pQCD mechanism, (ii) production is dominated by projectile scattering at large impact parameters, (iii) proportion of small $x_g$ contribution in the inclusive rate is approximately the same for $pp$ and $dAu$ collisions. 
 
 A lack of additional suppression of the $x_g \sim 0.01$ contribution to the double inclusive spectrum as compared to the suppression of the inclusive spectrum is explained in the post-selection mechanism as  due to a relatively small momentum of the produced gluon in the nucleus rest frame putting it far away from the BDR.
 
 It is difficult to reconcile  enumerated  features of the forward pion production data  with the 
 $2\to 1 $ mechanism \cite{Kharzeev} inspired by the color glass condensate model. In the scenario of  \cite{Kharzeev}  incoherent $2\to 2$ mechanism is neglected, a strong suppression of the recoil pion production is predicted. Also it   leads to a dominance of the central impact parameters and hence a larger multiplicity for the central hadron production in the events with the forward pion trigger. The observed experimental pattern indicates the models \cite{Dumitru:2005gt} which neglect contribution of the $2\to 2$ mechanism and consider only  $2\to 1$  processes strongly overestimates  inclusive cross section due to  the $2\to 1$ mechanism. 
 
Overall the observed regularities of inclusive forward pion production and forward central correlation  phenomenon give a strong indication of breakdown of the pQCD factorization due to the propagation of high energy partons through the nuclear media. The  modification of the nuclear gluon density  at small $ x < 0.01$ plays a  small role in this kinematics. 
 
 \section{Production of two forward pions and double-parton mechanism in $pp$ and $dA$ scattering}

 In Ref. \cite{GSV} we suggested that in order to study the effects of small $x$ gluon fields in the initial state one should   study production of two leading pions in nucleon - nucleus collisions. Recently the data were taken on production of two forward pions in $dAu$ collisions. The preliminary results of the studies of the reactions
 $pp \to \pi^0\pi^0+X,  d-Au \to \pi^0\pi^0+X$,  where one leading  pion served as a trigger and the second leading pion had somewhat smaller longitudinal and transverse momenta \cite{starqm09,phenixqm09}.
  The data indicate a strong suppression of the back to back production of pions in the central $dAu$ collisions. Also  a large fraction   of the double inclusive cross section  is isotropic in the azimuthal angle $\Delta\varphi$ 
of the two pions.  

 We performed a study  of this process in \cite{Strikman:2010bg}. We focused on explaining the isotropic component  of the double pion spectrum and 
 understand the origin of the suppression and in particular whether it is consistent with the post selection mechanism which we discussed above for the case of production of one forward pion.
  
  \subsection{Forward dipion production in $pp$ scattering}
  It is instructive to start with the case of $pp$ scattering. The leading twist contribution - $2\to 2$ mechanism - corresponds to the process in which a leading quark from the nucleon and a small x gluon from the target scatter to produce two jets with leading pions.  In this kinematics $x_g \le M^{2}(\pi \pi)/ x_q s_{NN}$. Production of two pions which together carry a large fraction of the nucleon momentum can occur only if $x_q$ is sufficiently close to one. 
  The results of our  calculation show that the  average value of $x_q$ for typical cuts of the RHIC experiments is pretty close to one.

Obviously it is more likely that two rather than one quark in a nucleon carry together $x $ close to one.
This suggests that in the   discussed  RHIC kinematics production the "double-scattering" contribution with two separate hard interactions in a single $pp$ collision could become important. Hence though  the discussed contribution is a "higher-twist", it is enhanced both by the probability of the relevant two quark configurations and the increase of the gluon density at small x which enters in the double-scattering in the second power.

One can derive the expression for the double-scattering  mechanism based on the analysis of the corresponding Feynman diagrams and express it through 
 the double  generalized parton densities   in the nucleons which we discussed in section 4. Similar to Eq.\ref{b1} 
 the cross section can be written in the form 
 \begin{eqnarray}
\frac{d^4\sigma}{dp_{T,1}d\eta_1dp_{T,2}d\eta_2} ={1\over \pi R^2_{int}}
\sum_{abcda'b'c'd'}
\int dx_a dx_b dz_c dx_{a'} dx_{b'} dz_{c'} \,\nonumber 
& & \\
 f_{aa'}^{H_1}(x_a, x_{a'})f_b^{H_2}(x_b)f_{b'}^{H_2}(x_{b'})D_c^{h_1}(z_c)\,
D_{c'}^{h_1}(z_{c'})\,
\frac{d^2 \hat{\sigma}^{ab\to cd}}{dp_{T,1} d\eta_1}\,
\frac{d^2 \hat{\sigma}^{a'b'\to c'd'}}{dp_{T,2} d\eta_2}\;
\label{mastermp}
\end{eqnarray}
Here  $f_{aa'}^{H_1}(x_a, x_{a'})$ is the double parton distribution. If the partons are not correlated, it is equal to the product of the single parton distributions.
For simplicity we neglected here correlations in the target as in our case $x's$ for gluons  are small. The  dimensional factor $ \pi R_{int}^2$ is given by Eq.\ref{b33}.
In our numerical calculations we used $\pi R_{int}^2 \approx 15 $ mb observed at the Tevatron which is smaller than the value obtained in the uncorrelated approximation 
(see discussion in section 4).

  We find that for the RHIC kinematics the only trivial correlation due to the fixed number of the valence quarks is important while the correlation between $x_a$ and $x_{a'}$ 	remains a small correction if we follow the quark counting rules to estimate the $x_{a'}$  dependence of $f_{a,a'}(x_{a},x_{a'}) $ for fixed $x_a$.
 The results of our calculation   indicate that 
 that the LT and double parton mechanisms are comparable for the kinematics of the RHIC experiments. This provides a natural explanation for the presence of a large component in the $pp \to \pi^0\pi^0+X$ cross section measured in 
 \cite{starqm09,phenixqm09} which does not depend on the azimuth angle 
 $\phi$. In fact the   number of events in the the pedestal is comparable to the number of events in the peak around $\phi \sim \pi$  which is dominated by the LT contribution indicating that the LT and double-parton contributions are indeed comparable (see Fig. 10a below).

Hence we conclude that the current experiments at RHIC have found a signal of double-parton interactions and that  future experiments at RHIC will be able to obtain a unique information about  double quark distributions in nucleons. It will be crucial for such studies to perform analyses for smaller bins of $\eta$ and preferably switch to the  analysis in bins of Feynman $x$.

 \subsection{Production of two forward pions and double-parton mechanism in $dAu$ scattering}
\begin{figure}[t]  
   \centering
   \includegraphics[width=0.8\textwidth]{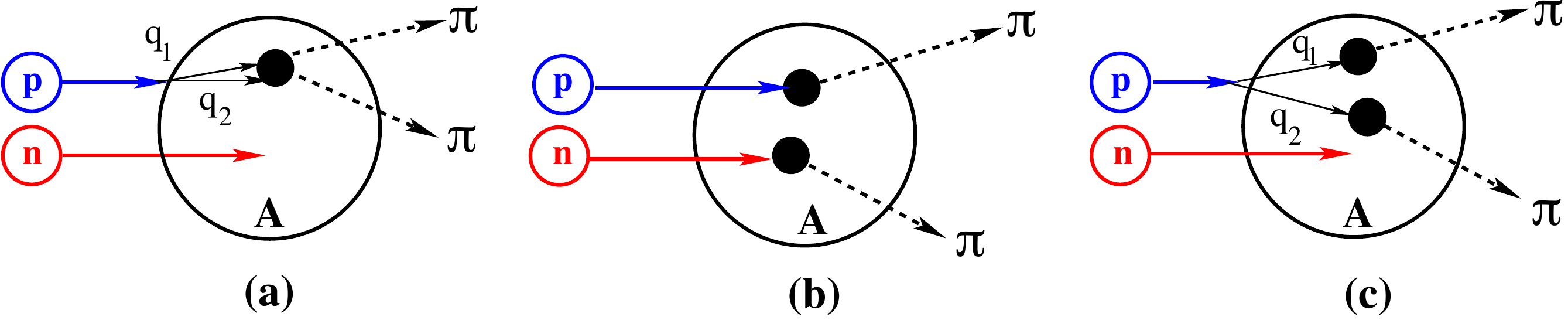} 
   \caption{Three  double parton mechanisms of dipion production.}
    \label{sketch}
 \end{figure}

Let us extend now  our results to the case of d-A scattering studied at RHIC.
In this case there are three distinctive double-parton mechanisms depicted in Fig.~ 
\ref{sketch}. The first two are the same as in the $pA$ scattering - scattering of two partons of the nucleon off two partons belonging to different nucleons (mechanism a), and off two   partons belonging to the same nucleon of the target (mechanism b) \cite{Strikman:2001gz}. The third mechanism, which is not present for $pA$ scattering is scattering of one parton of proton and one parton of the neutron off two partons of the nucleus.
Let us consider the ratio of the double-parton and leading twist contributions for $dA$ and $pp$ collisions 
\begin{equation}
r_{dA}= r_a+r_b+r_c= {\sigma_{DP}(dA)\over \sigma_{LT}(dA)}/{\sigma_{DP}(pp)\over \sigma_{LT}(pp)}.
\end{equation}
The contribution to $r_{dA}$ of the mechanisms (a), (c) is given by  
\cite{Strikman:2001gz}:
\begin{equation}
r_c= T(b)\sigma_{eff}; r_a = 1,
\end{equation}
where $T(b)$ is the standard nuclear profile function ($\int d^2bT(b)=A$).
Here we neglected nuclear gluon shadowing effect which is a small correction for the double-parton mechanism (cf. Ref.\cite{GSV}) but maybe important for the LT mechanism where $x_g $ maybe as low as $10^{-3}$ due to    the leading twist shadowing (see discussion below).
For the central $d-Au$ collisions $T_A\approx 2.2 fm^{-2}$  and so $r_a/r_c\sim 1/3$. The contribution (b) can be  calculated in a model independent way since  no parton correlations enter in this case. The ratio of $r_b$ and $r_c$ will be close to 1 at midrapidity, where correlations and valence-gluon scattering are not very important. Toward large rapidities, however, $r_b$ must become much larger than 
$r_c$, since it is not subject to the constraint $x_a+x_{a'} \le 1$ because of the fact that for (b) the proton and the neutron scatter independently.

As a result $r_{dAu}$ for small $b$ becomes of the order ten: $r_{dA}$ changes from $\sim$  9 to $\sim $ 12 for $\pi R_{int}^2 = 15 \div 20$ mb. 
 
\begin{figure}[t]  
   \centering
   \includegraphics[width=0.9\textwidth]{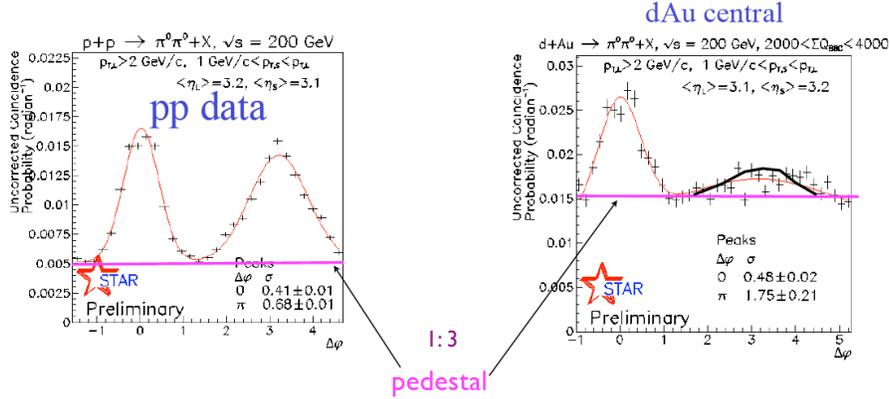} 
   \caption{STAR data for dipion production. Red curves are the Gaussian fit to the data. The horizontal magenta lines illustrate the strength of the double scattering while the solid black curve in the $dA$ plot illustrates the effect of the  reduction of the $2\to 2$ contribution by a factor of four as compared to the $pp$ case.}
    \label{topion}
 \end{figure}

 Since the single inclusive pion spectrum for $\eta_2 \sim 2 \div 3$ is suppressed by a factor of the order $R_A(b)= 1/3 \div 1/4$ we find for the ratio of the pedestals in $dAu$ and $pp$:
 \begin{equation}
R_{pedestal}= r_A R_A(b) \sim 2.5  \div 4,
\end{equation}
which should be compared with the experimental value of $R_{pedestal}\sim 3$. 
Hence we naturally explain the magnitude of the enhancement of the pedestal in central $dAu$ collision (see horizontal  magenta lines in Fig.\ref{topion} )

If  most of the pedestal in the kinematics studied at RHIC is due to the double-parton mechanism,  the uncertainties in the estimate of the rates due to this mechanism   and uncertainties in the strength of the suppression of the single inclusive forward pion spectrum at $b\sim 0$ would make it  very difficult to subtract this contribution with a precision necessary to find out whether all pedestal is due to double-parton  mechanism or there is a room for  a small contribution of the $2\to 1$ broadening mechanism  as it was assumed  in \cite{Albacete:2010pg}. Note also that in \cite{Albacete:2010pg} authors calculated  the ratio of the double inclusive cross section and the single inclusive cross section in the color glass condensate approach and compared this ratio with the data. However since  the single inclusive spectrum  is grossly overestimated  by the model (see discussion in section 8)  such procedure is not legitimate. 
 
 The suppression of the away peak originating from the LT contribution is  due to two effects: (i) the gluon shadowing 
  for $x\sim 10^{-3}$ and 
$b\le  3\, fm$ and $Q^2 \sim \mbox{few GeV}^2$ reduces the cross section by a factor of about two, (ii) stronger effect of effective fractional energy losses due to larger $x$ of the quark in the LT mechanism than in the double parton mechanism, leading to a suppression factor of the order two \cite{Strikman:2010bg}. All together this gives a suppression of the order of four as compared to the single pion trigger, which is consistent with the STAR observation - see solid curve in Fig.~\ref{topion}b.  It corresponds to the  overall suppression of the order of ten. This is pretty close to the low limit for the suppression estimated as 
the probability of 
  the "punch through" mechanism - contribution from the process where a quark scatters off one nucleon but does not encounter any extra nucleons at its impact parameter. Probability of such collisions at $b\sim 0$ for interaction with Au nucleus is of the order 
$5 \div 10\%$ \cite{Alvioli:2009ab}.

The data are consistent with suppression of the away peak by a factor $\ge 4$ and 
the reduction of the away peak relative to pedestal of the order of ten. The data may indicate that in addition to overall suppression there is some broadening of the away peak. Such effect is present  in the postselection mechanism, though for the very forward kinematics it is a correction to the effective energy losses.

For the LHC kinematics the discussed effects will be grossly amplified and extend to much wider range of $x$ - the same parton - target energy corresponds to rescaling of  x of the factor of $s_{LHC}/s_{RHIC} \ge 10^3$. In addition for $x\le 0.1$ 
the gluons give the dominant contribution while the BDR scale of $p_T^2$ is about a factor of two larger in this case (cf. Fig.\ref{bdrpt} ).

\section{Conclusions}
Studies aimed at understanding the underlying dynamics of $pp$ scattering at the LHC energies face a number of  challenges. The challenges discussed in this lecture include 
\begin{enumerate}
\item Building  models of inelastic collisions with realistic transverse parton distributions.

\item  Including effects of the   correlations between the partons in order to describe the rate of multiparton interactions. 

\item Realistic modeling of the BDR effects at moderate transverse momenta

\item Describing forward production at the LHC which is most sensitive to the BDR dynamics and in particular to the effect of effective fractional energy losses.
\end{enumerate}

\section*{Acknowledgements}
  
I thank my collaborators in the studies  which form the basis of  these lectures: B.Blok, Yu.Dokshitzer, L.Frankfurt, V.Guzey,  T.Rogers, W. Vogelsang, C.Weiss. 
My special thanks are to C.Weiss for critical reading of the first  draft of the manuscript.
The research was supported 
by the DOE grant No. DE-FG02-93ER40771. 

\end{document}